\documentclass[aps,prx,amsmath,superscriptaddress,twocolumn,bibnotes,longbibliography,floatfix]{revtex4-2}
\usepackage{amsmath}
\usepackage{amsfonts}
\usepackage{graphicx}
\usepackage{color}
\usepackage{dsfont}
\usepackage{verbatim}
\usepackage{mathtools}
\usepackage[normalem]{ulem}
\usepackage{comment}
\usepackage{stackrel}
\usepackage{bm}
\usepackage{booktabs}

\usepackage[colorlinks=true, 
linkcolor=blue, 
urlcolor=blue,
citecolor=blue]{hyperref}
\usepackage{orcidlink}
\usepackage[linesnumbered]{algorithm2e}

\RestyleAlgo{ruled}
\SetKwComment{Comment}{/* }{ */}
\definecolor{myblue}{rgb}{.93, .93, 1}

\newcommand{\bsub}{\begin{subequations}}
	\newcommand{\esub}{\end{subequations}}

\begin{document}
	
	\title{Symmetric localization of $\nu_{\text{tot}}=4/3$ fractional topological insulator edges}
	
	\author{Yang-Zhi~Chou~\orcidlink{0000-0001-7955-0918}}\email{yzchou@umd.edu}
	\affiliation{Condensed Matter Theory Center and Joint Quantum Institute, Department of Physics, University of Maryland, College Park, Maryland 20742, USA}

	\author{Sankar Das~Sarma}
	\affiliation{Condensed Matter Theory Center and Joint Quantum Institute, Department of Physics, University of Maryland, College Park, Maryland 20742, USA}
	
	\date{\today}
	
	\begin{abstract}
		Motivated by the recent twisted MoTe$_2$ experiment [arXiv:2601.18508], we develop a disordered interacting edge theory of a fractional topological insulator at $\nu_{\text{tot}}=4/3$, consisting of two time-reversal-conjugated $\nu=2/3$ fractional quantum Hall states. For an $S_z$-conserving edge, we uncover three distinct phases with two possible conductance values per edge in the long-edge limit: $\frac{2}{3}\frac{e^2}{h}$ and $\frac{4}{3}\frac{e^2}{h}$. In the presence of $S_z$-changing perturbations (e.g., Rashba spin-orbit coupling), an interaction-induced insulating edge state can emerge without breaking time-reversal or charge-conservation symmetry, corresponding to the absence of a topologically protected edge state. We show an exact mapping (with a special choice of parameters) to a noninteracting fermionic theory exhibiting Anderson localization, and the weak-coupling phase diagrams are also constructed, showing that symmetric localization can emerge regardless of other $S_z$-conserving perturbations. Our results showcase an explicit, experimentally relevant example that the edge-state two-terminal transport can yield false-negative results in identifying the $\nu_{\text{tot}}=4/3$ fractional topological insulators.
	\end{abstract}
	
	\maketitle
	
	\textit{Introduction. ---} Fractional topological insulators (FTIs) are strongly correlated gapped states that support fractionalized excitations while preserving time-reversal symmetry and charge conservation \cite{BernevigBA2006a,LevinM2009c,MaciejkoJ2010a}. (See reviews.~\cite{SternA2016,MaciejkoJ2015,NeupertT2015} and the references therein.) In two dimensions, the edge state of an FTI is described by fractionalized movers that form Kramers pairs, which can lead to a nontrivial two-terminal edge-state conductance.
	In addition, the edge states of two-dimensional (2D) FTI can be useful for engineering the parafermion zero modes \cite{ChengM2012,KlinovajaJ2014a}, a potential building block for topological quantum computations \cite{AliceaJ2016}. Despite extensive theoretical studies of FTIs, decisive evidence in the materials has remained elusive after two decades of exploration.

	During the past few years, the moir\'e transition metal dichalcogenides have become a promising platform for exploring the interplay between topology and correlation \cite{WuF2018,WuF2019b,PanH2020a,PanH2022,XieM2023b,MakKF2022,LiB2026}, owing to the tunable spin-valley-locked moir\'e valence bands with nonzero Chern numbers. For example, fractional quantum anomalous Hall effect has been observed in the twisted MoTe$_2$ with $\theta\approx 3.7^{\circ}$ \cite{CaiJ2023a,ZengY2023a,ParkH2023,XuF2023a,XuF2025b}, and a possible half-integer fractional quantum spin Hall effect was reported at hole doping $\nu_{\text{tot}}=3$ of a $\theta=2.1^{\circ}$ twisted MoTe$_2$ \cite{KangK2024c} ($\nu_{\text{tot}}$ is the number of carrier per moir\'e unit cell). The fractional quantum spin Hall experiment has motivated a number of theoretical studies \cite{ZhangYH2024b,JianCM2025,ZhangYH2024c,May-MannJ2025,SodemannVilladiegoI2024,ChouYZ2024,ReddyAP2024a,WenR2024a,KwanYH2026a,ChenF2024,XuC2025b,LuT2025}, but subsequent experiments found time-reversal breaking \cite{ParkH2024,XuF2025a,AnL2025,KangK2025b}, ruling out the time-reversal FTI possibility.  A very recent pump-probe circular dichroism experiment suggests a time-reversal symmetric (TRS) correlated state at $\nu_{\text{tot}}=4/3$ in twisted MoTe$_2$ \cite{WangY2026}, consistent with a 2D FTI made of a pair of time-reversal conjugated $\nu=2/3$ fractional quantum Hall (FQH) states -- a possibility predicted theoretically in Ref.~\cite{KwanYH2026a}. Nevertheless, the transport signature, which is a standard and important tool for identifying two-dimensional electronic topological phases, has not been experimentally verified.

\begin{figure}[t]
		\includegraphics[width=0.45\textwidth]{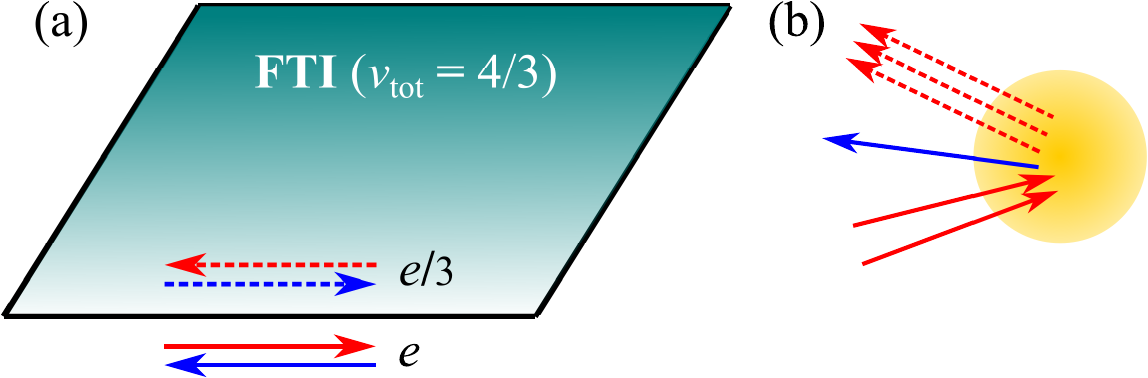}
		\caption{(a) The edge configuration of $\nu_{\text{tot}}=4/3$ FTI. The solid arrows indicate the charge-$e$ movers, and the dashed arrows indicate the charge-$e/3$ movers. The blue arrows can be viewed as the edge state of a spin-up $\nu=2/3$ FQH insulator; the red arrows can be viewed as the edge state of a spin-down $\nu=-2/3$ FQH insulator. (b) Illustration of the backscattering process $L^{\dagger}_{\uparrow}R_{\uparrow}L^{\dagger}_{\downarrow}R_{\uparrow}$ in $\mathcal{S}_{I,\text{loc}}$ [Eq.~(\ref{Eq:S_I_loc})]. Note that $L^{\dagger}_{\uparrow}$ is made of three $e/3$ quasiparticles (the three red dashed lines). The interaction in $\mathcal{S}_{I,\text{loc}}$ leads to a symmetric localized state in the strong coupling limit.
		}
		\label{Fig:Setup}
	\end{figure}

\begin{figure}[t]
		\includegraphics[width=0.4\textwidth]{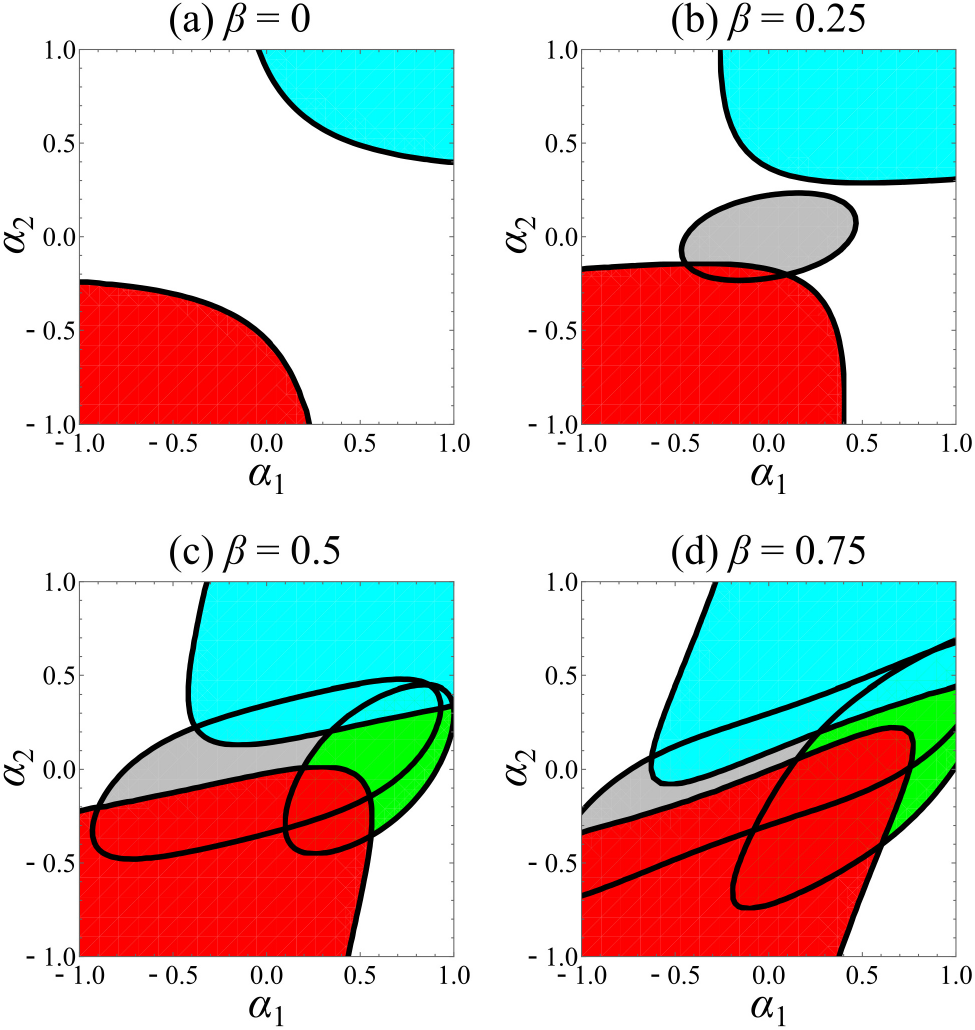}
		\caption{Weak-coupling phase diagram of the interacting FTI edge state. We select a few representative values of $\beta$ (encoding interactions between distinct Kramers pairs): (a) $\beta=0$, (b) $\beta=0.25$, (c) $\beta=0.5$, and (d) $\beta=0.75$. The cyan region indicates the $\mathcal{S}_{I,+}$ dominating phase; the red region indicate the $\mathcal{S}_{I,J}$ dominating phase; the green region is the symmetric localized phase; the $\mathcal{S}_{I,M}$ becomes the most relevant perturbation (but is unable to induce instability) in the gray region. The black curves are obtained by setting the scaling dimension equals to $2$ (for momentum-conserving $\mathcal{S}_{I,J}$ scattering) or $3/2$ (for other disorder scatterings). See main text and Appendix~\ref{App:PD} for detailed discussions.
		}
		\label{Fig:PD}
	\end{figure}

	In this work, we study the edge state of a 2D FTI at $\nu_{\text{tot}}=4/3$, made of a time-reversal pair of $\nu=2/3$ FQH states, as sketched in Fig.~\ref{Fig:Setup}(a). We analyze the possible phases driven by TRS perturbations and provide the corresponding two-terminal edge-state low-temperature conductance in the long-edge limit. See Table~\ref{Tab:phases} for a brief summary of the two-terminal low-temperature conductance and Fig.~\ref{Fig:PD} for the phase diagrams. Remarkably, we point out that an interaction-induced symmetric localized edge state can emerge from a generic edge state (i.e., without enforcing $S_z$ conservation) and provide an exact mapping (with a special choice of parameters) to a noninteracting fermionic theory manifesting Anderson localization. We further compute the phase diagram and confirm that such symmetric localization can emerge regardless of other perturbations. The responsible interaction backscattering is illustrated in Fig.~\ref{Fig:Setup}(b). Our results suggest that the two-terminal edge-state conductance may be insufficient to provide a decisive signature of $\nu_{\text{tot}}=4/3$ FTI, implying the need for novel characterizations for the search for FTIs.
	
	\begin{table}[t]
		\begin{ruledtabular}
			\begin{tabular}{c c c  c}
				Perturbation & $|\Delta S_z|$  & $\mathcal{N}$ & $G_{\infty}$ ($e^2/h$) \\
				\hline
				\hline
				None & N/A & 4 & 4/3 \\
				\hline
				$\mathcal{S}_{I,M}$ [Eq.~(\ref{Eq:S_I_M})] & 0 & 4 & 2/3\\
				$\mathcal{S}_{I,+}$ [Eq.~(\ref{Eq:S_I_+})] & 0 & 2 & 2/3\\
				$\mathcal{S}_{I,J}$ [Eq.~(\ref{Eq:S_I_J})] & 0 & 2 & 4/3\\
				\hline
				$\mathcal{S}_{I,\text{loc}}$ [Eq.~(\ref{Eq:S_I_loc})] & 1 & 0 & 0
			\end{tabular}
		\end{ruledtabular}
		\caption{Summary of different phases of FTI edge states. We list the properties of each phase driven by the associated perturbation in the strong-coupling limit. $|\Delta S_z|$ indicates the change of $S_z$ during the scattering, $\mathcal{N}$ denotes the number of low-energy modes, and $G_{\infty}$ indicates the two-terminal low-temperature conductance per edge in the infinite edge length limit.
		}
		\label{Tab:phases}
	\end{table}

	\textit{Edge theory as chiral bosons. ---} We consider a 2D product topological order at $\nu_{\text{tot}}=4/3$ consisting of two $\nu=2/3$ FQH states that form a time-reversal pair. Without loss of generality, we assume the spin-up electrons form a FQH state with a Hall conductivity $\sigma_H=(2/3)e^2/h$ and the spin-down electrons form a FQH state with a Hall conductivity $\sigma_H=-(2/3)e^2/h$ \cite{minTO}, with the edge state structure shown in Fig.~\ref{Fig:Setup}(a). Based on the earlier results of exact diagonalization \cite{WangY2026,KwanYH2026a}, this state is the ground state of the twisted MoTe$_2$ at hole doping $\nu_{\text{tot}}=4/3$. Using the chiral boson description \cite{WenXG1992,ChangAM2003,NeupertT2011a}, the imaginary-time action for the quadratic boson theory is given by 
	\begin{align}\label{Eq:S_0}
		\mathcal{S}_0=\frac{1}{4\pi}\int\limits_{\tau,x} \left(\partial_x\Phi^T\right)\left[\hat{\mathcal{K}}\left(i\partial_{\tau}\Phi\right)+\hat{\mathcal{V}}\left(\partial_x\Phi\right)\right],
	\end{align}
	where $\int\limits_{\tau,x}$ is a short-hand notation for $\int d\tau dx$, $\Phi$ is a four-component column vector made of real-valued chiral boson fields, $\hat{\mathcal{K}}$ is a $4\times 4$ K matrix encoding the commutation relations of the bosonic fields, and $\hat{\mathcal{V}}$ is a $4\times 4$ velocity matrix describing the velocities and non-universal forward scatterings among the chiral bosons. The chiral bosons obey the following commutation relation
	\begin{align}
		\left[\partial_x\Phi_a(x),\Phi_b(x')\right]=2\pi i \left(\hat{\mathcal{K}}^{-1}\right)_{ab}\delta(x-x').
	\end{align}
	
	For the FTI edge state considered in this work, the K matrix can be described by \cite{KaneCL1994,NeupertT2011a}
	\begin{align}\label{Eq:K_matrix}
		\hat{\mathcal{K}}=\begin{bmatrix}
			\hat{K} & 0\\
			0 & -\hat{K}
		\end{bmatrix},\,\,\,\hat{K}=\begin{bmatrix}
		1 & 0\\
		0 & -3
		\end{bmatrix}.
	\end{align}
	In the above expressions, the $\hat{\mathcal{K}}$ manifestly obeys time-reversal symmetry, and $\hat{K}$ is the K matrix for the $\nu=2/3$ FQH state, describing a charge-$e$ mover and a counter-propagating charge $e/3$ mover \cite{KaneCL1994}. The velocity matrix contains six independent variables and is given by
	\begin{align}
		\label{Eq:V_matrix}
			\hat{\mathcal{V}}=\begin{bmatrix}
				v_1 & \sqrt{3}u' & v_1' & \sqrt{3}u\\
				\sqrt{3}u' & 3v_2 & \sqrt{3}u & 3v_2'\\
				v_1' & \sqrt{3}u & v_1 & u'\\
				\sqrt{3}u & 3v_2' & \sqrt{3}u' & 3v_2
			\end{bmatrix},
	\end{align}
	where the factors of 3 and $\sqrt{3}$ are from the normalization convention.
	The corresponding integer-valued charge and the spin vectors given by $Q^T=[1,1,1,1]$ and $S^T=[1,1,-1,-1]$, respectively.
	The charge Hall conductivity is given by $\sigma_H=\nu_e\frac{e^2}{h}$, where $\nu_e=Q^T\hat{\mathcal{K}}^{-1}Q=0$; the spin Hall conductivity is given by $\sigma_{sH}=\nu_s\frac{e}{2\pi}$, where $\nu_s=\frac{1}{2}Q^T\hat{\mathcal{K}}^{-1}S=2/3$ \cite{NeupertT2011a}.  Note that the results of this work do not depend on the specific choice of the K matrix in Eq.~(\ref{Eq:K_matrix}), but the diagonal basis here is convenient for our analysis.
	
	In our theory, the four-component vector is $\Phi^T=[\varphi^R_{\uparrow},\varphi^L_{\uparrow},\varphi^L_{\downarrow},\varphi^R_{\downarrow}]$, where $\varphi^{\eta}_{s}$ is the chiral boson field with $\eta$ chirality ($R/L$) and spin $s$. Notably, $\varphi^R_\uparrow$ and $\varphi^L_\downarrow$ can be viewed as the $\nu=\pm 1$ integer quantum Hall edge states, while $\varphi^L_\uparrow$ and $\varphi^R_\downarrow$ are equivalent to the Laughlin quasiparticles of $\nu=\mp 1/3$. The electron field operator can be constructed by these chiral bosons, specifically, $R_{\uparrow}\sim e^{i\varphi^R_{\uparrow}}$, $L_{\uparrow}\sim e^{-i3\varphi^L_{\uparrow}}$, $R_{\downarrow}\sim e^{i3\varphi^R_{\downarrow}}$, and $L_{\downarrow}\sim e^{-i\varphi^L_{\downarrow}}$. Note that the minus signs in the exponents of the left-moving fermions are consistent with the total density operator given by $\rho_{\text{tot}}=\frac{1}{2\pi}Q^T\partial_x\Phi$. The factor of $3$ for $L_{\uparrow}$ and $R_{\downarrow}$ indicates that annihilating an electron is equivalent to the destruction of three Laughlin quasiparticles. 
	The time-reversal operation on chiral bosons are defined by \cite{NeupertT2011a} $(\varphi^R_{\uparrow},\varphi^L_{\uparrow})\rightarrow(\varphi^L_{\downarrow},\varphi^R_{\downarrow})$, $(\varphi^R_{\downarrow},\varphi^L_{\downarrow})\rightarrow(\varphi^L_{\uparrow}+\pi/3,\varphi^R_{\uparrow}+\pi)$, and $i\rightarrow -i$. One can show that the operations on fermions are given by $(R_{\uparrow},L_{\uparrow})\rightarrow (L_{\downarrow},R_{\downarrow})$, $(R_{\downarrow},L_{\downarrow})\rightarrow (-L_{\uparrow},-R_{\uparrow})$, and $i\rightarrow-i$.
	
	The quadratic boson theory is given by $\mathcal{S}_0$ [Eq.~(\ref{Eq:S_0})], describing a time-reversal pair of charge-$e$ movers ($\varphi^R_{\uparrow}$ and $\varphi^L_{\downarrow}$) and a time-reversal pair of charge-$e/3$ movers ($\varphi^L_{\uparrow}$ and $\varphi^R_{\downarrow}$). Tuning the matrix elements in $\hat{\mathcal{V}}$ of Eq.~(\ref{Eq:S_0}) does not induce a phase transition of the quadratic boson theory but modifies the scaling dimension of operators. In the absence of any backscattering perturbations, the two pairs of chiral bosons are not in equilibrium \cite{KaneCL1994}. As a result, the two-terminal conductance predicts $G=\frac{4}{3}\frac{e^2}{h}$ per edge state, independent of the velocity matrix $\hat{\mathcal{V}}$ \cite{ProtopopovIV2017,NosigliaC2018}. Backscattering interactions can remove low-energy modes and alter the transport properties, as studied in fractional quantum Hall edges \cite{MooreJE1998,MooreJE2002,VayrynenJI2022,YutushuiM2024a,ParkJ2024a}. Next, we study the phases driven by TRS backscattering perturbations and present the corresponding two-terminal conductance in the long-edge limit.

	\textit{$S_z$-conserving symmetry-allowed perturbations. ---} We investigate the symmetry-allowed backscattering perturbations (i.e., with time-reversal symmetry and charge conservation) with $S_z$ conservation. Note that $S_z$ conservation is \textit{not} an exact symmetry but an approximation. We focus only on the perturbations related to one-electron and two-electron scattering processes and ignore the higher-order terms (which are likely irrelevant or subleading). The main results are summarized in Table~\ref{Tab:phases}. We discuss each perturbation in detail below.

	First of all, we consider spin-conserving one-electron disorder backscattering described by
	\begin{subequations}\label{Eq:S_I_M}
		\begin{align}
			\mathcal{S}_{I,M}\!=&\!\int\limits_{\tau,x} \left[\eta_M(x)L^{\dagger}_{\uparrow}R_{\uparrow}+\eta_M^*(x)R^{\dagger}_{\downarrow}L_{\downarrow}+\text{H.c.}\right]\\
			=&\!\int\limits_{\tau,x} \left[\eta_M(x)e^{i(3\varphi^L_{\uparrow}+\varphi^R_{\uparrow})}+\eta_M^*(x)e^{-i(3\varphi^R_{\downarrow}+\varphi^L_{\downarrow})}+\text{H.c.}\right],
		\end{align}
	\end{subequations}
	where $\eta_M(x)$ is a complex-valued quenched disorder potential. For a zero-mean white-noise $\eta_M(x)$, $\mathcal{S}_{I,M}$ becomes a relevant perturbation when the scaling dimensions of $e^{i(3\varphi^L_{\uparrow}+\varphi^R_{\uparrow})}$ and $e^{-i(3\varphi^R_{\downarrow}+\varphi^L_{\downarrow})}$ are less than 3/2 \cite{GiamarchiT1988,GiamarchiT2003}. The disorder scattering process here is analogous to that in $\nu=2/3$ FQH edge states \cite{KaneCL1994,KaneCL1995,KaneCL1995a}, which is crucial for edge equilibration \cite{KaneCL1994,KaneCL1995,KaneCL1995a,ProtopopovIV2017,NosigliaC2018}. When the chiral bosons are equilibrated due to the disorder scatterings (either through elastic momentum relaxation \cite{KaneCL1994,NosigliaC2018} or inelastic scatterings \cite{ProtopopovIV2017}), the two-terminal conductance gives $G=\frac{2}{3}\frac{e^2}{h}$ per edge state in the long edge limit. Notably, these scatterings cannot remove any low-energy modes in the strong-coupling limit, as they do not satisfy the Haldane's instability criteria \cite{HaldaneFDM1995} (see Appendix~\ref{App:Haldane_criteria}). Thus, the phase driven by strong $\mathcal{S}_{I,M}$ is adiabatic to the ballistic phase dictated by the quadratic boson theory, although the conductance quanta are different in the long-edge limit.

	Besides the one-electron disorder scattering, we identify an $S_z$-conserving two-electron backscattering interaction, described by
	\begin{subequations}\label{Eq:S_I_+}
		\begin{align}
			\mathcal{S}_{I,+}=&\int\limits_{\tau,x}\left[\eta_+(x):L^{\dagger}_{\uparrow}L^{\dagger}_{\downarrow}R_{\downarrow}R_{\uparrow}:+\text{H.c.}\right]\\
			\sim&\int\limits_{\tau,x}\left[\eta_+(x)e^{i\left(\varphi^R_{\uparrow}+3\varphi^L_{\uparrow}+\varphi^L_{\downarrow}+3\varphi^R_{\downarrow}\right)}+\text{H.c.}\right]
		\end{align}
	\end{subequations}
	where $\eta_+(x)$ is a complex-valued quenched disorder potential and $:A:$ denotes the normal ordering of operator $A$. This interaction is analogous to the charge umklapp interaction in the spinful quantum wires \cite{GogolinAO2004,GiamarchiT2003} and mathematically equivalent to the $\mathcal{O}_+$ term in Ref.~\cite{ChouYZ2024}. For a zero-mean white-noise $\eta_+(x)$, $\mathcal{S}_{I,+}$ becomes relevant when the scaling dimension of $e^{i\left(\varphi^R_{\uparrow}+3\varphi^L_{\uparrow}+\varphi^L_{\downarrow}+3\varphi^R_{\downarrow}\right)}$ is less than 3/2 \cite{GiamarchiT1988,GiamarchiT2003}. The interaction here removes low-energy modes in the strong-coupling limit and induces a negative-drag correlation among the charge-$e$ and charge-$e/3$ channels. The negative-drag correlation can be inferred from the strong-coupling constraint, $\partial_{\tau}(\varphi^R_{\uparrow}+3\varphi^L_{\uparrow}+\varphi^L_{\downarrow}+3\varphi^R_{\downarrow})=0$, which is equivalent to $j_e+3j_{e/3}=0$ with $j_e$ ($j_{e/3}$) being the current in the charge-$e$ (charge-$e/3$) channel. Using the Ohm's law and $e^2/h$ for conductance of the charge-$e$ channel, one can straightforwardly show $G=\frac{2}{3}\frac{e^2}{h}$ per edge. Alternatively, a conductance formalism for fractionalized edge states \cite{ChouYZ2024} (which is a generalization of Ref.~\cite{OregY2014}) also yields the same result. The derived conductance here strictly holds in long-edge limit.
	
	Finally, a Josephson coupling between the charge-$e$ and the charge-$e/3$ channels \cite{ChouYZ2024} is also allowed by symmetry and is given by
	\begin{subequations}\label{Eq:S_I_J}
		\begin{align}
			\mathcal{S}_{I,J}=&J\int\limits_{\tau,x}\left[:R^{\dagger}_{\downarrow}L^{\dagger}_{\uparrow}L_{\downarrow}R_{\uparrow}+\text{H.c.}:\right]\\
			\sim&J\int\limits_{\tau,x}\left[e^{i\left(\varphi^R_{\uparrow}+3\varphi^L_{\uparrow}-\varphi^L_{\downarrow}-3\varphi^R_{\downarrow}\right)}+\text{H.c.}\right],
		\end{align}
	\end{subequations}
	where $J$ is the coupling constant for the Josephson coupling. Different from the previous two perturbations, this scattering process satisfies momentum conservation automatically, so the spatially uniform part of the interaction dominates. As a result, $\mathcal{S}_{I,J}$ becomes relevant when the scaling dimension of $e^{i\left(\varphi^R_{\uparrow}+3\varphi^L_{\uparrow}-\varphi^L_{\downarrow}-3\varphi^R_{\downarrow}\right)}$ is less than 2 (instead of 3/2 for the disordered cases) \cite{GiamarchiT2003,GogolinAO2004}. Notably, the interaction here is mathematically equivalent to the $\mathcal{O}_J$ term in Ref.~\cite{ChouYZ2024} and the interaction for the ``neutral-mode superconductivity'' among two counter-propagating $\nu=2/3$ FQH edges \cite{VayrynenJI2022}. The Josephson coupling $\mathcal{S}_{I,J}$ gaps out a pair of low-energy modes in the strong-coupling limit and then creates a positive-drag correlation among the charge-$e$ and charge-$e/3$ conducting channels, which can be shown explicitly. Assuming $J<0$, the strong-coupling constraint is given by $\varphi^R_{\uparrow}+3\varphi^L_{\uparrow}-\varphi^L_{\downarrow}-3\varphi^R_{\downarrow}=2\pi\times\text{integer}$, which can be recast into the current condition $j_e-3j_{e/3}=0$, corresponding to a two-terminal conductance $G=\frac{4}{3}\frac{e^2}{h}$ per edge state in the long-edge limit \cite{ChouYZ2024}.
	
	Finally, we note the three TRS perturbations are not compatible in the strong-coupling limit \cite{NeupertT2011a,MooreJE2002,XuC2006,ChouYZ2024}, so at most one perturbation manifests in the low-energy limit. (See Appendix~\ref{App:Haldane_criteria} for a discussion.)
	Therefore, the $S_z$-conserving edge state here allows for three distinct phases with two-terminal conductance values, $\frac{2}{3}\frac{e^2}{h}$ and $\frac{4}{3}\frac{e^2}{h}$ per edge, in the long edge limit.

	\textit{Symmetric localization in generic edge states. ---} Now, we discuss an interesting possibility due to the presence of $S_z$-changing scattering processes on the edge state. First, we focus on the $|\Delta S_z|=1$ TRS one-electron and two-electron backscattering perturbations, as they likely have larger coupling constants than other perturbations. The prominent one-electron $|\Delta S_z|=1$ backscattering is the random Rashba spin-orbit coupling \cite{XieHY2016}, which does not induce instability directly but can generate $S_z$-changing interactions.
	Among the leading $|\Delta S_z|=1$ perturbations, we identify an important two-electron interaction given by
	\begin{subequations}\label{Eq:S_I_loc}
		\begin{align}
			\mathcal{S}_{I,\text{loc}}=&\!\int\limits_{\tau,x}\!\left[\xi(x):L^{\dagger}_{\uparrow}R_{\uparrow}L^{\dagger}_{\downarrow}R_{\uparrow}:-\xi^*(x):R^{\dagger}_{\downarrow}L_{\downarrow}R^{\dagger}_{\uparrow}L_{\downarrow}:+\text{H.c.}\right]\\
			\sim&\!\int\limits_{\tau,x}\!\left[\begin{array}{c}
				\xi(x)e^{i(2\varphi^R_{\uparrow}+3\varphi^L_{\uparrow}+\varphi^L_{\downarrow})}\\[1mm]
				-\xi^*(x)e^{-i(\varphi^R_{\uparrow}+2\varphi^L_{\downarrow}+3\varphi^R_{\downarrow})}+\text{H.c.}
			\end{array}\right],
		\end{align}
	\end{subequations}
	where $\xi(x)$ is a complex-valued quenched disorder potential. For a zero-mean white-noise $\xi(x)$, $\mathcal{S}_{I,\text{loc}}$ becomes relevant when the scaling dimensions of $e^{i(2\varphi^R_{\uparrow}+3\varphi^L_{\uparrow}+\varphi^L_{\downarrow})}$ and $e^{-i(\varphi^R_{\uparrow}+2\varphi^L_{\downarrow}+3\varphi^R_{\downarrow})}$ are less than 3/2 \cite{GiamarchiT1988,GiamarchiT2003}. In the strong-coupling limit, $e^{i(2\varphi^R_{\uparrow}+3\varphi^L_{\uparrow}+\varphi^L_{\downarrow})}$ and $e^{-i(\varphi^R_{\uparrow}+2\varphi^L_{\downarrow}+3\varphi^R_{\downarrow})}$ operators can simultaneously remove all the chiral bosons from the low-energy theory, resulting in an insulating state in the long edge limit. (See a discussion in Appendix~\ref{App:Haldane_criteria}.) Remarkably, such an insulating state preserves time-reversal symmetry and charge conservation, i.e., the absence of topological protection on the boundary of an FTI.

	An intuitive ways to examine the symmetry property is to express the interaction $\mathcal{S}_{I,\text{loc}}$ in terms of a one-fermion backscattering term. First, we identify $R_{\uparrow}L^{\dagger}_{\downarrow}R_{\uparrow}$ and $L_{\downarrow}R^{\dagger}_{\uparrow}L_{\downarrow}$ as two composite fermionic charge-$e$ operators, denoted by $\mathcal{R}_1$ and $\mathcal{L}_1$, respectively. This representation is reminiscent of the $\nu=1/3$ FQH fermionic operators in the coupled-wire construction \cite{KaneCL2002,TeoJCY2014}. Notably, $\mathcal{R}_1$ and $\mathcal{L}_1$ form a Kramers pair, similar to the regular helical edge state. The interaction given by $\mathcal{S}_{I,\text{loc}}$ can be viewed as deflecting one fermion between two different Kramers pairs (e.g., $L^{\dagger}_{\uparrow}R_{\uparrow}L^{\dagger}_{\downarrow}R_{\uparrow}\rightarrow L^{\dagger}_{\downarrow}\mathcal{R}_1$), which can induce localization without breaking time-reversal symmetry or charge conservation \cite{HasanMZ2010a,QiXL2011a}. In Appendix~\ref{App:Mapping}, we show an exact mapping to a noninteracting fermionic model that manifests the symmetric localization with a special choice of parameters. The symmetry property discussed here applies generally regardless of the noninteracting fermion representation.

	The absence of topological protected edge state has been pointed out by Levin and Stern \cite{LevinM2009c,LevinM2012,SternA2016} for the FTI with an even $\nu_{s}/e^*$  (2 in our case), where $\nu_s$ is the dimensionless spin Hall conductivity [see discussion below Eq.~(\ref{Eq:K_matrix})] and $e^*$ is the minimal charge (in the unit of $e$). However, explicit constructions are rare, and the backscattering perturbations may be in higher-order processes, without a transparent physical consequence for the realistic materials. Our results provide an explicit construction, showing how the symmetric insulating edge state arises in a minimal model that is relevant to the recent twisted MoTe$_2$ experiment \cite{WangY2026}. 
	
	\textit{Weak-coupling TRS phase diagram. ---} With the possible interaction-driven phases discussed above, we construct the edge-state phase diagrams. Following the general treatment in Refs.~\cite{MooreJE1998,XuC2006}, we compute the scaling dimensions of each perturbation and construct the weak-coupling phase diagrams. We note that the spontaneous time-reversal breaking phases \cite{XuC2006,WuC2006,ChouYZ2024,ChouYZ2018,ChouYZ2019,KainarisN2014}, which are driven by $|\Delta S_z|=2$ scatterings (presumably parametrically small), are neglected in this work. The detailed derivations are provided in Appendix~\ref{App:PD}, and we briefly summarize the main ideas and results below. 
	
	First, the scaling dimensions depend only on three independent parameters \cite{XuC2006} [$\alpha_1$, $\alpha_2$, and $\beta$ defined in Eq.~(\ref{Eq:M_B_matrices})], instead of six parameters in the velocity matrix [Eq.~(\ref{Eq:V_matrix})]. These parameters capture the interactions within the Kramers pair ($\alpha_1$ and $\alpha_2$) as well as between two distinct Kramers pairs ($\beta$). Positive (negative) values indicate repulsive (attractive) forward-scattering interactions, and $\alpha_1=\alpha_2=\beta=0$ corresponds to the absence of any forward scattering interactions (i.e., velocity matrix $\hat{\mathcal{V}}$ is purely diagonal). The general phase diagram can be constructed with these three independent parameters.
	
	Then, we compute the scaling dimensions of all the TRS perturbations discussed in this work. The momentum-conserving scattering $\mathcal{S}_{I,J}$ becomes relevant when the associated scaling dimension is less than $2$ \cite{GiamarchiT2003,GogolinAO2004}. Other perturbations are disorder scatterings, and the critical value is $3/2$ \cite{GiamarchiT1988,GiamarchiT2003}. In addition, multiple perturbations can become relevant simultaneously, and no two of the perturbations are compatible (see Appendix~\ref{App:Haldane_criteria} for a discussion). As we discussed previously, the $\mathcal{S}_{I,M}$ term cannot induce instability, so we assume it is always the weakest perturbation. We further assume a hierarchy in the interaction strength such that the perturbations with $\Delta S_z=0$ are stronger than the $\mathcal{S}_{I,\text{loc}}$ (with $|\Delta S_z|=1$) \cite{ChouYZ2024}.

	In Fig.~\ref{Fig:PD}, we construct the weak-coupling phase diagrams as functions of $\alpha_1$ and $\alpha_2$ with $\beta\ge 0$. The phases driven by the $\mathcal{S}_{I,+}$ (cyan region) and $\mathcal{S}_{I,J}$ (red region) expand when $\beta$ is increased. The $\mathcal{S}_{I,M}$ dominating region (in gray) manifests in Figs.~\ref{Fig:PD}(b)-(d) and is adiabatic to the ballistic phase (white region). The phase driven by $\mathcal{S}_{I,\text{loc}}$ (green region) emerges in Figs.~\ref{Fig:PD}(c)-(d), suggesting that the symmetric localization can be realized regardless of other $\Delta S_z=0$ perturbations. For $\beta<0$, the phase diagram is similar to the $\beta=0$ case [Fig.~\ref{Fig:PD}(a)], but the phases driven by the $\mathcal{S}_{I,+}$ and $\mathcal{S}_{I,J}$ are smaller.

	\textit{Discussion. ---} In this work, we investigate the edge-state two-terminal conductance of a $\nu_{\text{tot}}=4/3$ FTI, relevant to the recent twisted MoTe$_2$ experiment \cite{WangY2026}. For edge states with strict $S_z$ conservation, we find that the edge theory remains conducting for the leading $S_z$-conserving TRS perturbations. However, the two-terminal conductance is nonuniversal and depends on the details. In the long edge limit, we expect the two-terminal edge-state conductance converges to either $\frac{2}{3}\frac{e^2}{h}$ (dominant $\mathcal{S}_{I,M}$ or $\mathcal{S}_{I,+}$) or $\frac{4}{3}\frac{e^2}{h}$ (dominant $\mathcal{S}_{I,J}$) per edge, demonstrating the topological protection in the $S_z$-conserving edge states. 
	
	Meanwhile, $S_z$ conservation is not exact, and the edge states have lower symmetry (e.g., absence of inversion symmetry). As a result, terms like Rashba spin-orbit coupling and ''Rashba-dressed'' interactions can become significant on the edge states despite the fact that $S_z$ conservation approximately holds in the bulk. For generic edge states without $S_z$ conservation, the $\mathcal{S}_{I,\text{loc}}$ can result in an insulating state without breaking time-reversal symmetry or charge conservation. The low-temperature conductance is given by $G\sim e^{-L/\xi_c}$ for $L\gg \xi_c$, where $\xi_c$ is the localization length. At small finite temperatures (assuming the localized order is not completely melted), we expect insulator-like behavior, i.e., the resistance increases as temperature is lowered. Thus, the generic edge state may not be sharply distinguished from the trivial Anderson localized edge state through the standard transport measurement. Other notable $S_z$-changing perturbations are $|\Delta S_z|\ge 2$ processes (e.g., $:L^{\dagger}_{\downarrow}R_{\uparrow}R^{\dagger}_{\downarrow}L_{\uparrow}:$) and inelastic interactions \cite{SchmidtTL2012,LezmyN2012,ChouYZ2015,KainarisN2014,KainarisN2017,ZhangX2021}. However, the $|\Delta S_z|\ge 2$ perturbations are likely parametrically small as the approximate $S_z$ conservation disfavors larger $|\Delta S_z|$ scatterings, and these inelastic interactions \cite{SchmidtTL2012,LezmyN2012,ChouYZ2015,KainarisN2014,KainarisN2017,ZhangX2021} cannot induce instability to the quadratic edge theory. Thus, we conclude that the symmetric localization due to $\mathcal{S}_{I,\text{loc}}$ is the dominant correction to the $S_z$-conserving edge.

	We conclude by discussing several interesting future directions. 
	In this work, we assume the edge state can be described by one pair of charge-$e$ movers and one pair of charge-$e/3$ movers. However, this is not the only edge state configuration, as the boundary may go through edge reconstruction, similar to the $\nu=2/3$ FQH edge state \cite{WangJ2013,NosigliaC2018}. It is interesting to investigate the reconstructed edge theory (which contains more than two Kramers pairs) and repeat the analysis of this work. The symmetric localization is not unique to the edge states of $\nu_{\text{tot}}=4/3$ FTI studied in this work. It will be useful to systematically investigate the edge-state phases and conductance values in other filling fractions, e.g., $\nu_{\text{tot}}=4/5$.
	Finally, it is desirable to develop novel characterizations of the FTI beyond edge-state transport (e.g., momentum-resolved tunneling spectroscopy \cite{HuberM2005}, visualizing edge states \cite{YuJ2025}, and shot noise measurements \cite{TikhonovES2015}), especially since our work provides a sharp, experimentally pertinent example that an FTI can have symmetric localized edge states.
	
	\begin{acknowledgments}
		Acknowledgments-- Y.-Z.C. thank Seth Musser and Yahui Zhang for useful conversations.
		This work is supported by the Laboratory for Physical Sciences.
	\end{acknowledgments}

	%\bibliography{fTI}
	

	\newpage
	
	%\vspace{0.5cm}
	
	\begin{center}
		\textbf{\Large End Matter}
	\end{center}
	
	\appendix

	\section{Instability criteria and compatibility conditions}\label{App:Haldane_criteria}
	
	Here, we examine the TRS perturbations with Haldane's instability criteria and the compatibility conditions \cite{HaldaneFDM1995,MooreJE2002,XuC2006}. We first provide the scattering vectors for all backscattering perturbations considered in this work and then compute the instability and compatibility conditions.
	
	In the bosonic theory, a backscattering of fermions can be described by a vertex operator
	\begin{align}
		\mathcal{O}_{\mathcal{X}}(\tau,x)\sim \exp\left[im_{\mathcal{X}}^T\Phi(\tau,x)\right],
	\end{align}
	where $m_{\mathcal{X}}$ is a four-component integer vector encoding the scattering. For the one-electron disorder scattering $\mathcal{S}_{I,M}$ [Eq.~(\ref{Eq:S_I_M})], the scatter vectors correspond to $m_M^T=[1,3,0,0]$ and $m_M'^T=[0,0,1,3]$. The two-electron backscattering $\mathcal{S}_{I,+}$ [Eq.~(\ref{Eq:S_I_+})] corresponds to an integer-valued scattering vector $m^T_+=[1,3,1,3]$. The Josephson coupling $\mathcal{S}_{I,J}$ [Eq.~(\ref{Eq:S_I_J})] corresponds to an integer-valued scattering vector $m^T_J=[1,3,-1,-3]$. The $|\Delta S_z|=1$ two-electron interaction $\mathcal{S}_{I,\text{loc}}$ [Eq.~(\ref{Eq:S_I_loc})] correspond to scattering vectors $m_{\text{loc}}^T=[2,3,1,0]$ and $m_{\text{loc}}'^T=[1,0,2,3]$.
	
	Now, we discuss Haldane's instability criteria and the compatibility of different perturbations \cite{HaldaneFDM1995,MooreJE2002,XuC2006}. For a scattering process described by an integer-valued scattering vector $m_{\mathcal{X}}$, the scaling dimension $\Delta(m_{\mathcal{X}})$ is bound by \cite{MooreJE2002,XuC2006,NeupertT2011a,NeupertT2014}
	\begin{align}\label{Eq:D_m}
		\Delta(m_{\mathcal{X}})\ge\left|\frac{m_{\mathcal{X}}^T\hat{\mathcal{K}}^{-1}m_{\mathcal{X}}}{2}\right|\equiv \mathcal{D}(m_{\mathcal{X}}).
	\end{align}
	When $\mathcal{D}(m_{\mathcal{X}})=0$, the vertex operator $e^{im_{\mathcal{X}}^T\Phi}$ can remove a pair of low-energy modes (e.g., a right-moving mode and a left-moving mode) in the strong-coupling fixed point \cite{HaldaneFDM1995,MooreJE2002,XuC2006,NeupertT2011a,NeupertT2014}. The above situation is Haldane's instability criteria. Using Eq.~(\ref{Eq:D_m}), we obtain
	\begin{subequations}
		\begin{align}
			&\mathcal{D}(m_{M})=\mathcal{D}(m_{M}')=1,\\
			&\mathcal{D}(m_{+})=0,\\
			&\mathcal{D}(m_{J})=0,\\
			&\mathcal{D}(m_{\text{loc}})=\mathcal{D}(m_{\text{loc}'})=0.
		\end{align}
	\end{subequations}
	The results above suggest that $\mathcal{S}_{I,M}$ cannot remove low energy modes in the strong-coupling limit, while other perturbations ($\mathcal{S}_{I,+}$, $\mathcal{S}_{I,J}$, and $\mathcal{S}_{I,\text{loc}}$) satisfy the Haldane's instability criteria.
	
	Another important question is whether two distinct perturbations are compatible. Two vertex operators, $e^{im_{1}^T\Phi}$ and $e^{im_{2}^T\Phi}$, can coexist in the strong-coupling limit if $\mathcal{C}(m_1,m_2)=m_1^T\hat{\mathcal{K}}^{-1}m_2$. We find that $\mathcal{C}(m_M,m_M')=\mathcal{C}(m_{\text{loc}},m_{\text{loc}}')=0$ and other pairs are nonzero. Since $\mathcal{D}(m_{\text{loc}})=\mathcal{D}(m_{\text{loc}'})=0$ and $\mathcal{M}(m_{\text{loc}},m_{\text{loc}}')=0$, the two backscattering interactions in $\mathcal{S}_{I,\text{loc}}$ can simultaneously remove two pairs of the low-energy modes, resulting in an insulating edge state in the infinite length limit, as discussed in the main text.
	
	\section{Mapping to fermionic theory}\label{App:Mapping}
	
	In this Appendix, we provide an exact mapping to a noninteracting fermion theory in a special limit.
	First, we set $u=u'=0$ in the velocity matrix [Eq.~(\ref{Eq:V_matrix})]. In this case,
	$v_1>0$ and $v_2>0$ encodes the velocity of the chiral bosons, $v_1'$ and $v_2'$ described the forward scatterings among different movers. $v_1',v_2'>0$ denotes the repulsive interaction, and the stability of the bosonic theory requires that $|v_1'|<v_1$ and $|v_2'|<v_2$.
	The above choice ignores the coupling between the charge-$e$ and the charge-$e/3$ channels, making them manifestly decoupled. Then, we express the chiral boson fields by the phase-like and the density-like bosonic field via $\varphi^R_{\uparrow}=\phi_1+\theta_1$, $\varphi^L_{\downarrow}=-\phi_1+\theta_1$, $\varphi^R_{\downarrow}=(\phi_2+\theta_2)/3$, and $\varphi^L_{\uparrow}=(-\phi_2+\theta_2)/3$. Using these new bosonic variables, one can straightforwardly show that the Hamiltonian (corresponding to $\mathcal{S}_0+\mathcal{S}_{I,\text{loc}}$) is expressed by $\hat{H}_1+\hat{H}_2+\hat{H}_{12}$, where
	\begin{subequations}
		\begin{align}
			\label{Eq:H_1}\hat{H}_1=&\frac{u_1}{2\pi}\int dx\left[K_1\left(\partial_x\phi_1\right)^2+\frac{1}{K_1}\left(\partial_x\theta_1\right)^2\right],\\
			\label{Eq:H_2}\hat{H}_2=&\frac{\bar{u}_2}{2\pi}\int dx\left[K_2\left(\partial_x\phi_2\right)^2+\frac{1}{K_2}\left(\partial_x\theta_2\right)^2\right],\\
			\label{Eq:H_12}\hat{H}_{12}=&\frac{1}{(2\pi\alpha)^2}\int dx\left[\begin{array}{c}
				\xi(x)e^{i(\phi_1+3\theta_1-\phi_2+\theta_2)}\\[1mm]
				-\xi^*(x)e^{i(\phi_1-3\theta_1-\phi_2-\theta_2)}+\text{H.c.}
			\end{array}\right].
		\end{align}
	\end{subequations}
	In $\hat{H}_1$ and $\hat{H}_2$, we have introduced the velocities $u_1=\sqrt{(v_1-v_1')(v_1+v_1')}$, $\bar{u}_2=\sqrt{(v_2-v_2')(v_2+v_2')}/3$ and the Luttinger parameters $K_1=\sqrt{(v_1-v_1')/(v_1+v_1')}$, $K_2=\sqrt{(v_2-v_2')/(v_2+v_2')}$. Note that $K_1=1$ ($K_2=1$) corresponds to $v_1'=0$ ($v_2'=0$). In Eq.~(\ref{Eq:H_12}), we have restored the factor $1/(2\pi\alpha)^2$ (with $\alpha$ being the ultraviolet length scale) based on the fermion-boson identity in the standard Abelian bosonization \cite{GiamarchiT2003,GogolinAO2004}.
	
	In a special limit, the bosonic Hamiltonian $\hat{H}_1+\hat{H}_2+\hat{H}_{12}$ allows for a noninteracting fermionic description. To show this, we introduce the following fermionic fields,
	\begin{align}
		\mathcal{R}_1=&e^{i(\phi_1+\tilde{\theta}_1)}/\sqrt{2\pi\alpha},\,\,\,\mathcal{L}_1=e^{i(\phi_1-\tilde{\theta}_1)}/\sqrt{2\pi\alpha},\\
		\mathcal{R}_2=&e^{i(\phi_2+\theta_2)}/\sqrt{2\pi\alpha},\,\,\,\mathcal{L}_2=e^{i(\phi_2-\theta_2)}/\sqrt{2\pi\alpha},
	\end{align}
	where $\tilde{\theta}_1=3\theta_1$, and $\alpha$ is an ultraviolet length scale. The Klein factors have been ignored as they are not essential to our analysis. Under the time-reversal operation, $\mathcal{R}_n\rightarrow-\mathcal{L}_n$ and $\mathcal{L}_n\rightarrow\mathcal{R}_n$, for $n=1,2$, similar to the helical edge state of 2D TRS topological insulators \cite{XuC2006,WuC2006,KaneCL2005a,HasanMZ2010a,QiXL2011a}.

	Now, we discuss the fermionic representation of the Hamiltonian $\hat{H}_1+\hat{H}_2+\hat{H}_{12}$. Setting $K_1=1/3$, the $\hat{H}_1$ term [Eq.~(\ref{Eq:H_1})] becomes \cite{GiamarchiT2003,GogolinAO2004,ChouYZ2023b}
	\begin{align}
		\hat{H}_1=&\frac{u_1/3}{2\pi}\int dx\left[\left(\partial_x\phi_1\right)^2+\left(\partial_x\tilde\theta_1\right)^2\right]\\
		\rightarrow &\bar{u}_1\int dx\left[\mathcal{R}_1^{\dagger}\left(-i\partial_x\mathcal{R}_1\right)-\mathcal{L}_1^{\dagger}\left(-i\partial_x\mathcal{L}_1\right)\right],
	\end{align}
	where $\bar{u}_1=u_1/3$. Similarly, $\hat{H}_2$ [Eq.~(\ref{Eq:H_2})] with $K_2=1$ becomes
	\begin{align}
		\hat{H}_2\rightarrow \bar{u}_2\int dx\left[\mathcal{R}_2^{\dagger}\left(-i\partial_x\mathcal{R}_2\right)-\mathcal{L}_2^{\dagger}\left(-i\partial_x\mathcal{L}_2\right)\right].
	\end{align}
	$\hat{H}_1$ and $\hat{H}_2$ describes 1D fermions with linear dispersion. 
	Finally, the $\hat{H}_{12}$ can be expressed by
	\begin{align}
		\hat{H}_{12}\rightarrow\int dx \left[\bar\xi(x)\mathcal{L}_2^{\dagger}\mathcal{R}_1-\bar\xi^*(x)\mathcal{R}_2^{\dagger}\mathcal{L}_1+\text{H.c.}\right],
	\end{align}
	where $\bar\xi(x)=\xi(x)/(2\pi\alpha)$.
	Notably, $\hat{H}_{12}$ represents a single-particle disorder backscattering in the fermionic basis here.
	Thus, $\hat{H}_1+\hat{H}_2+\hat{H}_{12}$ is equivalent to the two copies of TRS topological insulator edge states with TRS backscattering disorder, and Anderson localization without symmetry breaking is anticipated \cite{HasanMZ2010a,QiXL2011a}. 
	
	The above mapping relies on the specific form of the velocity matrix $\hat{\mathcal{V}}$ with $v_1'=4v_1/5$, $v_2'=0$, and $u=u'=0$ (i.e., $K_1=1/3$ and $K_2=1$). For a general velocity matrix, the $\hat{H}_1$ and $\hat{H}_2$ gain forward-scattering interactions, making the fermionic theory not exactly solvable. We expect that the symmetric localization survives in the presence of small but finite interactions. The precise phase boundary requires a separate calculation, which we do not pursue here.
	
	\section{Derivation of TRS phase diagram}\label{App:PD}

	Here, we construct the phase diagram following the general treatment in Refs.~\cite{MooreJE1998,XuC2006}. The velocity matrix in Eq.~(\ref{Eq:V_matrix}) can be expressed by
	\begin{align}
		\hat{\mathcal{V}}=\hat{M}\hat{B}_1\hat{B}_2\hat{R}\begin{bmatrix}
			v_1^{(0)} & 0 & 0 & 0\\
			0 & v_2^{(0)} & 0 & 0\\
			0 & 0 & v_1^{(0)} & 0\\
			0 & 0 & 0 & v_2^{(0)}
		\end{bmatrix}
		\hat{R}^T\hat{B}_2\hat{B}_1\hat{M},
	\end{align}
	where
	\begin{subequations}\label{Eq:M_B_matrices}
		\begin{align}
		\hat{M}=&\begin{bmatrix}
			1 & 0 & 0 & 0\\
			0 & \sqrt{3} & 0 & 0\\
			0 & 0 & 1 & 0\\
			0 & 0 & 0 & \sqrt{3}
		\end{bmatrix},\\
		\hat{B}_1=&\begin{bmatrix}
		\cosh(\alpha_1) & 0 & \sinh(\alpha_1) & 0\\
		0 & \cosh(\alpha_2) & 0 & \sinh(\alpha_2)\\
		\sinh(\alpha_1) & 0 & \cosh(\alpha_1) & 0\\
		0 & \sinh(\alpha_2) & 0 & \cosh(\alpha_2)
		\end{bmatrix},\\
		\hat{B}_2=&\begin{bmatrix}
			\cosh(\beta) & \sinh(\beta) & 0 & 0\\
			\sinh(\beta) & \cosh(\beta) & 0 & 0\\
			0 & 0 & \cosh(\beta) & \sinh(\beta)\\
			0 & 0 & \sinh(\beta) & \cosh(\beta)
		\end{bmatrix},\\
		\hat{R}=&\begin{bmatrix}
			\cos(\delta) & 0 & 0 & -\sin(\delta)\\
			0 & \cos(\delta) & -\sin(\delta) & 0\\
			0 & \sin(\delta) & \cos(\delta) & 0\\
			\sin(\delta) & 0 & 0 & \cos(\delta)
		\end{bmatrix}.
	\end{align}
	\end{subequations}
	In this representation of $\hat{\mathcal{V}}$, there are two velocities ($v_1^{(0)}$ and $v_2^{(0)}$), three ``boost'' parameters ($\alpha_1$, $\alpha_2$, and $\beta$), and one rotation parameter ($\delta$). With this parameterization, the scaling dimension of a vertex operator is given by
	\begin{align}\label{Eq:Scaling_dim_gen}
		\Delta(m)=m^T\hat{\Delta}m,
	\end{align}
	where $m$ is the integer vector characterizing the scattering process and
	\begin{align}\label{Eq:Delta_matrix}
		\hat{\Delta}=\frac{1}{2}\hat{M}^{-1}\hat{B}_1^{-1}\hat{B}_2^{-1}\hat{B}_2^{-1}\hat{B}_1^{-1}\hat{M}^{-1}.
	\end{align}
	In the expression of Eq.~(\ref{Eq:Delta_matrix}), the velocities and the rotation parameter do not affect the scaling dimension. Thus, the general phase diagram in our case can be described by three independent boost parameters, $\alpha_1$, $\alpha_2$, and $\beta$. We compute the scaling dimensions of the operators in $\mathcal{S}_{I,M}$, $\mathcal{S}_{I,+}$, $\mathcal{S}_{I,J}$, and $\mathcal{S}_{I,\text{loc}}$, corresponding to $m_M^T=[1,3,0,0]$, $m_+^T=[1,3,1,3]$, $m^T_J=[1,3,-1,-3]$, and $m^T_{\text{loc}}=[2,3,1,0]$, respectively. Then, we identify the region such that $\Delta(m)<2$ (momentum-conserving perturbation) \cite{GiamarchiT2003,GogolinAO2004} for $m=m_J$ and $\Delta(m)<3/2$ (disorder perturbation) \cite{GiamarchiT1988,GiamarchiT2003} for other scattering processes. The results are plotted in Fig.~\ref{Fig:PD}.

\end{document}